\documentclass[aps,prl,twocolumn,superscriptaddress,english]{revtex4}

\usepackage{graphicx}
\usepackage{dcolumn}
\usepackage{braket}
\usepackage{amsmath} 
\usepackage{amsmath, amssymb, amsfonts}
\usepackage[sectionbib]{bibunits}
\defaultbibliography{apssamp}
\defaultbibliographystyle{apsrev}

\begin{document}
 \bibliographyunit[\section]

\preprint{APS/123-QED}

\title{Exploring the Berezinskii-Kosterlitz-Thouless Transition \\ in a Two-dimensional Dipolar Bose Gas}


\author{Yifei He}
\thanks{These authors contributed to this work equally.}
\affiliation{Department of Physics, The Hong Kong University of Science and Technology, Clear Water Bay, Kowloon, Hong Kong, China}

\author{Ziting Chen}
\thanks{These authors contributed to this work equally.}
\affiliation{Department of Physics, The Hong Kong University of Science and Technology, Clear Water Bay, Kowloon, Hong Kong, China}

\author{Haoting Zhen}
\affiliation{Department of Physics, The Hong Kong University of Science and Technology, Clear Water Bay, Kowloon, Hong Kong, China}

\author{Mingchen Huang}
\affiliation{Department of Physics, The Hong Kong University of Science and Technology, Clear Water Bay, Kowloon, Hong Kong, China}

\author{Mithilesh K Parit}
\affiliation{Department of Physics, The Hong Kong University of Science and Technology, Clear Water Bay, Kowloon, Hong Kong, China}

\author{Gyu-Boong Jo}
\email{gbjo@ust.hk}
\affiliation{Department of Physics, The Hong Kong University of Science and Technology, Clear Water Bay, Kowloon, Hong Kong, China}
\affiliation{IAS Center for Quantum Technologies, The Hong Kong University of Science and Technology, Kowloon, Hong Kong, China}

\date{\today}

\begin{abstract}
Long-range and anisotropic dipolar interactions induce complex order in quantum systems. It becomes particularly interesting in two-dimension (2D), where the superfluidity with quasi-long-range order emerges via Berezinskii-Kosterlitz-Thouless (BKT) mechanism, which still remains elusive with dipolar interactions. Here, we observe the BKT transition from a normal gas to the superfluid phase in a quasi-2D dipolar Bose gas of erbium atoms. Controlling the orientation of dipoles, we characterize the transition point by monitoring extended coherence and measuring the equation of state. This allows us to gain a systematic understanding of the BKT transition based on an effective short-range description of dipolar interaction in 2D. Additionally, we observe anisotropic density fluctuations and non-local effects in the superfluid regime, which establishes the dipolar nature of the 2D superfluid. Our results lay the ground for understanding the behavior of dipolar bosons in 2D and open up opportunities for examining complex orders in a dipolar superfluid.
\end{abstract}

\maketitle


\section*{Introduction}

\begin{figure*}
\includegraphics[scale=0.52]{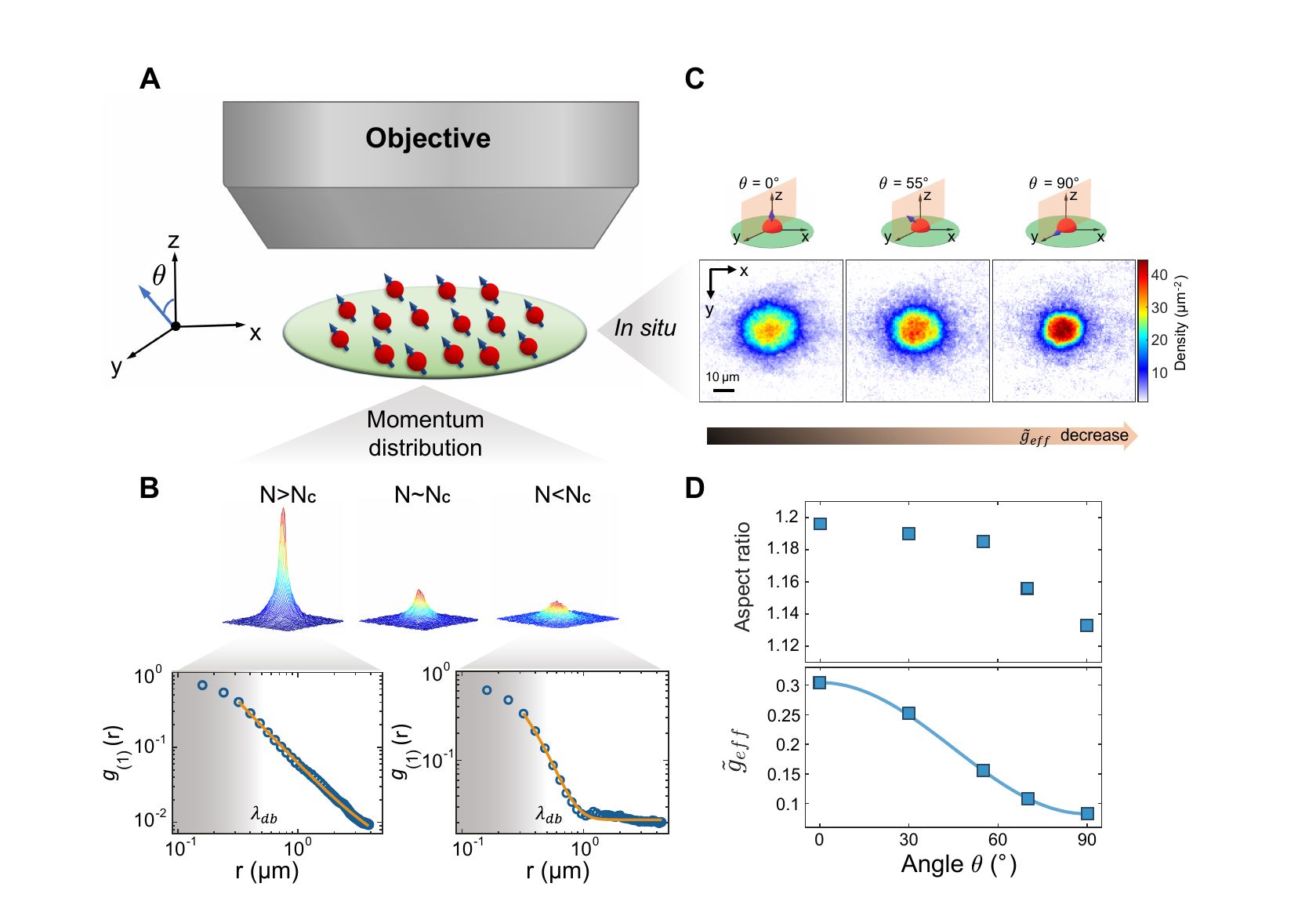}
\caption{\textbf{Dipolar 2D Bose gas with tunable dipole angle}\ \textbf{(A)} Schematics of the experiment. Atoms are loaded into a 2D trap and atomic dipoles are polarized by a bias magnetic field at angle $\theta$ from $z$ axis in the $y-z$ plane. \textbf{(B)} Probing the BKT transition of dipolar 2D sample in momentum space. When the sample crosses the BKT transition point a sharp zero momentum peak appears and the trap averaged first-order correlation $g_{1}(r)$ shows algebraic decay instead of exponential decay. The blue empty circles are $g_{1}(r)$ obtained by the Fourier transform of momentum distribution normalized by $g_{1}(0)=1$ and the orange curve on the left(right) is the power-law(exponential) fitting starting from the distance around $0.7\lambda_{db}$. \textbf{(C)} \textit{in situ} density distribution of sample with different $\theta$, showing we can tune $\widetilde{g}_{eff}$ by tuning the dipole orientation. \textbf{(D)} Upper: Aspect ratio change when changing the dipole orientation due to the anisotropic nature of DDI. Lower: Theoretical $\widetilde{g}_{eff}$ as a function of angle $\theta$ in our experimental parameter. Filled squares above~(below) are measured~(theoretical) values for 5 specific angles($\theta=0^{\circ},30^{\circ},55^{\circ},70^{\circ},90^{\circ}$) where we conduct experiment.}
\end{figure*}

Quasi-two-dimensional (quasi-2D) dipolar system is a long-sought platform where numerous novel phenomena rising from the anisotropic long-range dipole-dipole interaction (DDI) are predicted, including stable 2D bright soliton\cite{pedri2005two}, anisotropic superfluidity~\cite{ticknor2011anisotropic} and coherence~\cite{ticknor2012anisotropic},
long-range vortex-vortex interactions~\cite{mulkerin2013anisotropic} and anisotropic number fluctuation~\cite{baillie2014number}. In oblate trap geometry, various many-body phenomena induced by DDI such as anisotropic superfluidity~\cite{wenzel2018anisotropic}, supersolidity~\cite{norcia2021two} and 2D roton excitation~\cite{schmidt2021roton} have been observed with ultracold magnetic atoms like dysprosium and erbium. Nevertheless, none of these works fulfill the kinematically 2D condition in the quasi-2D regime. Although dipolar atoms~\cite{koch2008stabilization,pasquiou2010control,du2023atomic} and molecules~\cite{li2023tunable} have been arranged in 2D arrays in an optical lattice, the nature of 2D superfluid has not yet been addressed. More recently, a novel quantum solid has been observed in a 2D lattice system with dipolar excitons and dipolar erbium atoms in the extended Hubbard model~\cite{lagoin2022extended,su2023dipolar}.

In contrast to the Bose-Einstein condensation (BEC) driven by quantum statistics, the interaction-driven Berezinskii-Kosterlitz-Thouless (BKT) transition~\cite{berezinskii1972destruction,kosterlitz1973ordering} connects the normal gas phase and superfluid phase in 2D~\cite{hadzibabic2006berezinskii,desbuquois2012superfluid}. The BKT transition has been extensively explored in contact interacting quasi-2D Bose and Fermi gases, including scale-invariant behavior and universality around critical point~\cite{hadzibabic2006berezinskii,hung2011observation,yefsah2011exploring,desbuquois2012superfluid,desbuquois2014determination,murthy2015observation}. Moreover, the interaction-dependent BKT transition was  predicted~\cite{prokof2001critical,holzmann2010universal} and measured in both position~\cite{hung2011observation} and momentum space~\cite{fletcher2015connecting} by tuning s-wave scattering length via Feshbach resonance. Since interaction plays an essential role in BKT transition, how DDI affects the BKT transition remains a particularly interesting open question~\cite{defenu2023long}. In recent experiments with excitons~\cite{dang2019defect,dang2020observation}, however, the only interaction considered is the DDI, which is approximated as a purely repulsive isotropic short-range contact interaction~\cite{filinov2010berezinskii} likewise a 2D contact-interacting gas. Furthermore, the dipole orientation is not adjustable, which prevents the emergence of novel features that could be induced by the long-range and anisotropic nature of DDI.

In this work, we produce ultracold dipolar $^{168}$Er atoms in the quasi-2D harmonic trap and examine the BKT transition with controllable DDI by dipole orientation together with tunable s-wave scattering contact interaction. Possessing a large magnetic moment up to $\mu_m=7~\mu_B$, $\mu_B$ being Bohr magneton, a gas of Erbium atoms is suitable for examining the BKT scenario with tunable dipoles~\cite{chomaz2022dipolar}. The experiments are conducted under a 600~mG bias magnetic field and the corresponding s-wave scattering length $a_s$ is 140~$a_0$~\cite{patscheider2022determination}, where $a_0$ is the Bohr radius. To characterize the strength of DDI, the dipolar scattering length is defined as $a_{dd}=\mu_0\mu_m^2m/12\pi\hbar^2$, where $\mu_0$ is vacuum permeability and $m$ is the atomic mass. Here, $a_{dd}$ is given as $66~a_0$ for erbium atoms. 

Although DDI is a long-range interaction in 3D, it forms a uniform local interaction potential like contact interaction under homogeneous quasi-2D condition with mean field solution $\mu=n[g_s+g_{dd}(3\cos^2(\theta)-1)]$~\cite{ticknor2011anisotropic,baillie2014number}, where $\mu$ is chemical potential, $\theta$ is the angle between the dipole orientation and normal vector of 2D plane, and $n$ is the 2D atomic density. The quasi-2D contact and DDI coupling constant are defined as  $g_s=\sqrt{8\pi}\hbar^2 a_s/ml_z$ and $g_{dd}=\sqrt{8\pi}\hbar^2a_{dd}/ml_z$, respectively, with the axial harmonic oscillator length of $l_z=\sqrt{\hbar/m\omega_z}$. We introduce effective 2D coupling strength $g_{eff}(\theta)=g_s+g_{dd}(3\cos^2(\theta)-1)$ and dimensionless form $\widetilde{g}_{eff}(\theta) = \sqrt{8\pi}[a_s+a_{dd}(3\cos^2(\theta)-1)]/l_z$ in quasi-2D dipolar gases where the contribution from DDI can be tuned from repulsive to attractive by varying the dipole orientation $\theta$.

\subsection*{Experiments}

To create a 2D trap geometry, we achieve tight confinement in the axial direction using a focused red-detuned 532~nm optical sheet beam, while the radial confinement is provided by an vertical optical dipole trap (vODT) together with the sheet beam, resulting a trap frequency $(\omega_x,\omega_y,\omega_z)=2\pi\times(14.3, 17.0, 1070) $ Hz. Throughout the experiments, we first prepare a nearly pure BEC in the Zeeman sublevel $m_J=-6$ $(J=6)$~\cite{seo2020efficient,seo2023apparatus}, and then adiabatically load it into our quasi-2D trap in 300~ms while keeping the magnetic field direction along $z$. In the following, we adiabatically rotate the magnetic field from the $z$ direction to the target angle $\theta$ in $y-z$ plane in 100~ms to tune the dipole orientation. We rotate the magnetic field very slowly to avoid exciting any collective mode. After dipoles are prepared to the target angle $\theta$ as shown in Fig.~1A, we wait for another 500~ms for the system to equilibrate. In the 2D sample, the kinematically 2D condition is fulfilled with $\mu<\hbar\omega_z$ and $k_BT\lesssim\hbar\omega_z$, where $k_B$ is the Boltzmann constant, $T$ is the temperature and $\hbar$ is reduced Planck constant. Therefore, atoms predominantly lie in the ground state of the harmonic oscillator in $z$ direction and the axial motion is frozen. For a typical sample, $\sim 94\%$ atoms lie in axial ground state.

To characterize the BKT transition point, we employ the momentum focusing technique~\cite{tung2010observation,murthy2014matter,fletcher2015connecting} to probe the in-plane momentum distribution $n(\textbf{k})$ of the 2D sample. When the atom number exceeds the critical point $N_c$ of the BKT transition, a sharp peak appears at zero momentum which signals the extended coherence over the thermal DeBroglie wavelength $\lambda_{db}=h/\sqrt{2\pi mk_BT}$ in the cloud as shown in Fig.~1B~\cite{fletcher2015connecting}. In addition, direct evidence of the BKT transition in the dipolar 2D sample can be found in the first-order correlation function $g_1(r)$ obtained by 2D Fourier transform of momentum distribution~\cite{murthy2015observation}. Crossing the superfluid transition, $g_1(r)$ shows a clear algebraic decay in an intermediate range $1~\lambda_{db} \sim 8~\lambda_{db}$ while in normal phase $g_1(r)$ decays exponentially to background value within the order of $1 \lambda_{db}$. Different background values come from normalization process.

To further understand the microscopic properties of the dipolar 2D sample, we take \textit{in-situ} images of the cloud from a high-resolution image system (N.A.=0.28 and spatial resolution $\sim 1 \mu m$). When changing the orientation of the dipole from out-of-plane ($\theta=0^{\circ}$) to in-plane ($\theta=90^{\circ}$), we observe a decrease in the cloud size and an increase in the center density, as shown in Fig.~1C. This is consistent with the fact that the interaction strength $\widetilde{g}_{eff}$ is a function of the angle $\theta$.  Changing $\theta=0^{\circ}$ to $90^{\circ}$ corresponds to $\widetilde{g}_{eff}=0.3\sim0.08$ as shown in Fig.~1D, which is well within the weakly interacting regime $\widetilde{g}_{eff}<1$~\cite{hung2011observation,fletcher2015connecting}. To be noticed, this parameter regime is stable and away from any phonon-instability and roton-instability~\cite{santos2003roton,ronen2007radial,blakie2012roton,mishra2016dipolar}. In addition, when dipoles are tilted in-plane, one may expect the aspect ratio to change due to the magnetostriction effect~\cite{stuhler2005observation}. 
Although DDI is reduced to a local isotropic interaction $g_{dd}(3\cos^2(\theta)-1)$ in the homogeneous 2D system, we still observe a minimal difference in terms of aspect ratio between $\theta=0^{\circ}$ and $90^{\circ}$ in the 2D harmonic trap at around $6\%$ as shown in Fig.~1D. This change closely agrees with the extended Gross-Pitaevskii equation~(eGPE) simulation in mean-field regime~\cite{cai2010mean}. 

\subsection*{Determination of $N_c^{BKT}$ for the BKT transition}

\begin{figure}[!h]
\begin{center}
\includegraphics[scale=0.52]{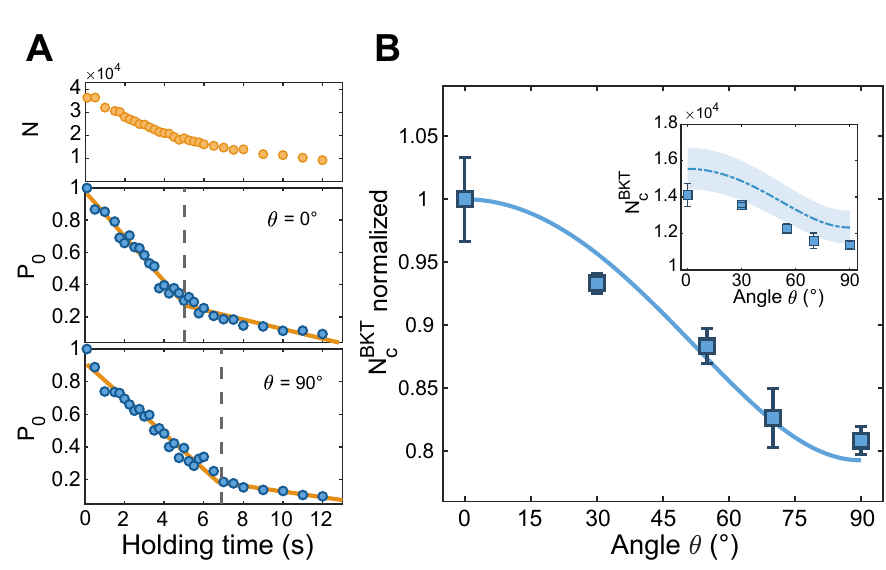}\\
\caption{\textbf{Measuring critical atom numbers of BKT transition with different dipole angles.}\ \textbf{(A)} Examples of the time evolution of atom number(upper panel) and zero momentum peak(middle and bottom panels) of 2D samples which represent extended coherence. An empirical piecewise linear fit~(orange solid line) is used to determine the transition point $t_c$~(vertical dash line). \textbf{(B)} Model-independent measurement of $N_c^{BKT}$ normalized by the critical number at $\theta=0^{\circ}$, blue solid curve is theoretical scaling factor from eq.(1) based on $\widetilde{g}_{eff}$. Inset: Critical atom number only considering atoms the ground state of axial harmonic oscillator. The dash-dotted curve is $N_c^{BKT}$ calculated by eq.(1) based on temperature fitted by critical samples and independently measured trap frequency. Blue shaded area denotes systematic uncertainty. Error bars are statistical errors.} \label{fig2}
\end{center}
\end{figure}

Non-interacting ideal Bose gas can go through a BEC transition at finite temperature in 2D harmonic potential because the atoms in excited states will saturate at $N_c^0=\frac{\pi^2}{6}(\frac{k_BT}{\hbar\omega_r})^2$. However, when interaction exists, the BEC phase transition will be replaced with the interaction-driven BKT transition. The critical atom number of BKT transition is calculated based on classical field simulation~\cite{holzmann2010universal} as
\begin{equation}
    N_c^{BKT}(\widetilde{g}) \approx N_c^0(1+3\frac{\widetilde{g}}{\pi^3}\ln^2{(\frac{\widetilde{g}}{16})}+\frac{6\widetilde{g}}{16\pi^2}[15+\ln{(\frac{\widetilde{g}}{16})}])
\end{equation}
where $\widetilde{g}$ is the dimensionless 2D coupling strength. This result has been verified by measuring the critical atom number under different interaction strengths with different $a_s$~\cite{fletcher2015connecting}.

To determine the critical point in a dipolar Bose gas, we take model-independent measurements by holding the sample in the trap for variable time to control atom number and monitor the evolution of zero momentum peak $P_0$ of the 2D sample~\cite{fletcher2015connecting}. While the atom number varies smoothly as a function of holding time (Fig.~2A top panel), the evolution of zero momentum peak shows two distinct regions. We use a piecewise linear function to extract the onset of coherence and fit the transition holding time $t_c$ empirically. We count the total atom number of sample at $t_c$ to determine the raw value of $N_c^{BKT}$, showing the clear shift when the dipole angle $\theta$ varies (Fig.~2A middle and bottom panel). Similar transition point can also be observed from first order correlation function $g_1(r)$\cite{SI}.

The shift of the critical atom number is quantitatively elucidated in $N_c^{BKT}$ normalized by the critical atom number measured at $\theta=0^\circ$, as a function of $\theta$ in Fig.~2B. Since all the sequences are the same before adiabatically rotating the magnetic field, it is not expected for the temperature to exhibit systematic differences for samples with different angles, regardless of any model. The axial ground state population is mainly affected by temperature while only modified by interaction. Therefore, we provide a fully model-independent result on the scaling factor of $N_c^{BKT}$ as a function of angle by showing the relative change of $N_c^{BKT}$ normalized by the raw $N_c^{BKT}$ measured at $\theta=0^\circ$. It suggests the scaling factor of $N_c^{BKT}$ in the 2D harmonic trap is determined by $\widetilde{g}_{eff}(\theta)$ following $N_c^{BKT}(\widetilde{g}_{eff})$ which only includes the short range contribution from DDI. 

We also take the fitting based on Hartree-Fock mean field theory with $g_{eff}$ on high momentum tail to extract the temperature and axial ground state population of our critical samples~\cite{fletcher2015connecting,SI}. The absolute $N_c^{BKT}$ only considering the atoms in the axial ground state are plotted in the inset. For the $N_c^{BKT}$ measurement, the temperature is kept relatively high, resulting in approximately 76\% of the atoms lying in the axial ground state. The blue dash-dotted curve is the calculated $N_c^{BKT}$ based on the fitting temperature from momentum focusing image of critical samples. The absolute $N_c^{BKT}$ is about 10\% below the theory prediction which is possibly due to the systematic calibration error or imperfect thermometry\cite{SI}. The beyond Born approximation correction on $a_{dd}$~\cite{oldziejewski2016properties} and the uncertainty on $a_s$\cite{patscheider2022determination} are estimated to be a few percent level in total, which only cause less than 1\% correction on predicted $N_c^{BKT}$.

\subsection*{Equations of state (EoS) measurement}

\begin{figure}
\includegraphics[scale=0.53]{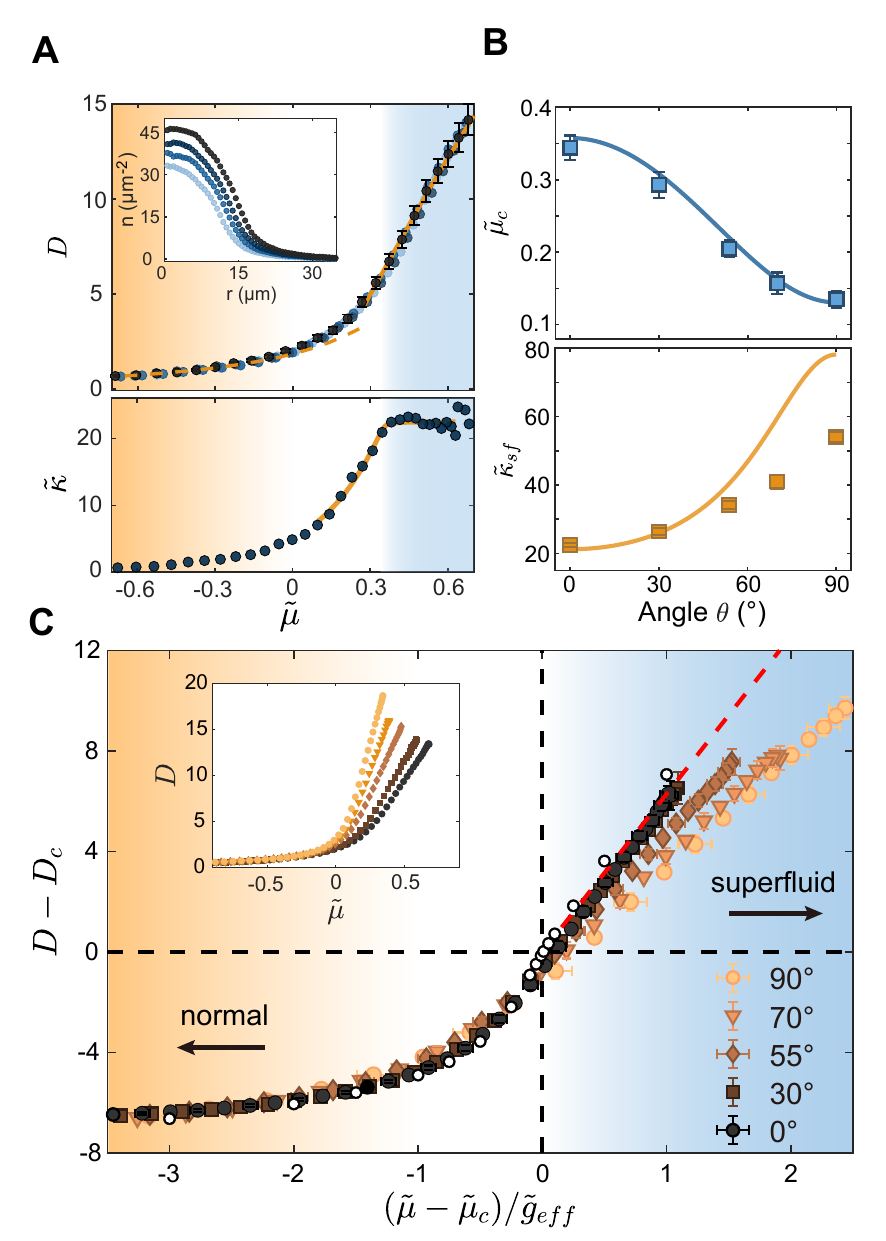}\\
\caption{\textbf{Equations of state measurement.}\ \textbf{(A)} \textbf{Upper panel}: Scale-invariance behavior of quasi-2D dipolar samples at $\theta=0^{\circ}$ viewed by EoS constructed from different samples. The EoS follows the HFMF prediction (dash line) in normal regime (shaded orange) and classical field prediction (solid lind) in superfluid regime (shaded blue) like purely contact gas~\cite{hung2011observation}. Inset: Corresponding radial density profile with different atom number and temperature. \textbf{Lower panel}: Scaled-compressibility calculated from EoS. The orange line is the empirical fitting to determine $\widetilde{\mu}_c$ and $\widetilde{\kappa}_{sf}$. \textbf{(B)} Critical chemical potential $\mu_c$ and compressibility in superfluid regime $\widetilde{\kappa}_{sf}$ as a function of angle $\theta$. The solid line above is $\mu_c$ predicted by classical field theory with $\widetilde{g}_{eff}$. The solid line below is the compressibility estimated by TF approximation. \textbf{(C)} Rescaled EoS around BKT critical point of different angle $\theta$. The open circles are Monte Carlo calculations from~\cite{prokof2002two}. The slope of red dash line represents TF limit. Inset: Original EoS. Error bars denote statistical standard error and fitting error.} \label{fig1}
\end{figure}

After probing the global coherence properties by monitoring the zero momentum peak, we investigate the local properties of 2D dipolar samples in position space by taking \textit{in situ} images. Here, we measure the \textit{in situ} density distribution for 5 specific angle $\theta=0^{\circ},30^{\circ},55^{\circ},70^{\circ},90^{\circ}$ and axial ground state population is kept larger than $90\%$. Based on local density approximation (LDA) $\mu(r)=\mu_0-1/2m\omega_r^2r^2$, we azimuthally average the density along equi-potential line which represents the equal local chemical potential to obtain $n(r)\ vs.\ \mu(r)$. When dipoles are oriented along the $z$ direction, the interaction within the plane is fully isotropic and the density distribution follows the shape of the trap. However, for samples with tilted dipoles, the density distribution no longer perfectly follows the trap geometry due to anisotropic DDI. Nevertheless, in our most anisotropic sample ($\theta=90^\circ$), the deviation between the equi-potential curve and the equi-density curve is only $\pm 3\%$ of the sample radius, which is below the optical resolution limit of our system. Therefore, averaging based on the trap geometry does not introduce significant uncertainty for our anisotropic sample~\cite{SI}.

The finite temperature simulation suggests that the thermal tail of a dipolar quasi-2D gas has the standard profile for a 2D contact gas~\cite{ticknor2012finite}. Therefore we take the standard thermometry based on Hartree-Fock mean field~(HFMF) theory~\cite{tung2010observation,hung2011observation,yefsah2011exploring,hadzibabic2008trapped} with $\widetilde{g}_{eff}$ to extract the global chemical potential $\mu_0$, the temperature $T$ and the density profile in axial harmonic ground state $n_0$ from averaged image to construct the equations of state~(EoS) for ground state atoms~\cite{SI}. We first test the scale invariance in the dipolar 2D sample by producing samples with atom number from $18,000$ to $35,000$ and corresponding temperature from $30~nK$ to $44~nK$, then measure the EoS. The EoS $ D\ vs.\ \widetilde{\mu}$ of samples with dipoles along $z$ direction($\theta=0^\circ$) is plotted in Fig.~3A upper panel, where $\widetilde{\mu}=\mu/k_BT$ is the scaled chemical potential and $D=n_0\lambda_{db}^2$ is the 2D phase space density~(PSD). The result suggests that the scale invariance still exists in weakly interacting dipolar 2D Bose gases that PSD $D$ is only the function of scaled chemical potential $\widetilde{\mu}$. For a dipolar 2D gas with dipoles perpendicular to plane, the EoS behaves the same as a contact interacting Bose gas~\cite{hung2011observation} that the normal phase follows the HFMF prediction\cite{holzmann2010universal} while the superfluid phase follows the classical field prediction $D=2\pi\widetilde{\mu}/\widetilde{g}_{eff}+\ln{\left(2D\widetilde{g}_{eff}/\pi-2\widetilde{\mu}\right)}$~\cite{prokof2002two}. The scale-invariant behavior is also observed in the EoS of 2D dipolar gases with tilted dipoles although it cannot be described by $\widetilde{g}_{eff}$ in the superfluid regime due to non-local effect~\cite{SI}.

To identify a superfluid transition from the density profile, we compute the scaled compressibility $\widetilde{\kappa}=\partial{D}/\partial{\widetilde{\mu}}$ from the derivative of EoS (Fig.~3A lower panel). The compressibility first increases with the increase of $\widetilde{\mu}$ in the normal gas regime, and quickly approaches a constant value defined by Thomas-Fermi~(TF) limit (for example $\widetilde{\mu}=\widetilde{g}_{eff}D/2\pi$ for $\theta=0^{\circ}$) when it crosses the transition\cite{holzmann2008kosterlitz}. The same transition feature has also been observed in the density fluctuation due to fluctuation-dissipation theorem\cite{hung2011observation}. We empirically fit the transition feature of the scaled compressibility to estimate the critical chemical potential $\widetilde{\mu}_c$ for different angles, together with the compressibility in superfluid regime $\widetilde{\kappa}_{sf}$ (Fig.~3A lower panel, also see Methods). The result of critical chemical potential shows good agreement with the theoretical prediction~\cite{prokof2001critical} $\widetilde{\mu}_c=\widetilde{g}/\pi\ln{\left(13.2/\widetilde{g}\right)}$ with $\widetilde{g}=\widetilde{g}_{eff}(\theta)$ as a function of angle as shown in Fig.~3B upper panel. This result is consistent with our model-independent $N_c$ measurement in momentum space that the local interaction $\widetilde{g}_{eff}$ determines the superfluid phase transition point in a dipolar 2D system.  When $\theta<55^{\circ}$, $\widetilde{\kappa}_{sf}$ closely follows the TF limit $2\pi/\widetilde{g}_{eff}$~(solid line in Fig.~3B lower panel), while when dipoles are tilted predominantly in plane, a deviation from TF limit occurs due to the non-local term calculated in mean-field regime~\cite{cai2010mean} as lower panel of Fig.~3B shows.

 Universality around BKT transition point was predicted~\cite{prokof2002two} and observed~\cite{hung2011observation} in weakly interacting contact 2D Bose gas, $D-D_c=H\left(X\right),\ X=\frac{\widetilde{\mu}-\widetilde{\mu}_c}{\widetilde{g}}$ where $H$ is a generic function. We also test the universality of our dipolar 2D sample around the BKT transition point. Because the result in Fig.~3B already shows good agreement with theoretical prediction, we directly take~\cite{prokof2001critical} 

\begin{equation}
    \widetilde{\mu}_c=\frac{\widetilde{g}_{eff}}{\pi}\ln{\left(\frac{13.2}{\widetilde{g}_{eff}}\right)}, \ D_c=\ln{\left(\frac{380}{\widetilde{g}_{eff}}\right)}
\end{equation}
to plot $D-D_c\ vs.\ X $ in Fig.~3C. 
It turns out that the universality of the dipolar 2D system survives in $-3<X<0$ from normal gas phase covering the fluctuation region until the phase transition point. When dipoles predominantly aligns along $z$ axis~($\theta=0^\circ,30^\circ$), universal behavior further extend to superfluid regime as a purely contact gas~\cite{hung2011observation,prokof2002two}. 
The suppression of density when $\theta$ is larger than $55^\circ$ comparing with the TF limit in the superfluid region can be described by the eGPE simulation result~\cite{cai2010mean}. Generally speaking, the non-local term of DDI in a 2D harmonic trapping condensate has two effects, one is the anisotropy which changes the aspect ratio of the cloud when dipoles are tilted, and the other is to shift the overall density in the trap, both of them are significantly enhanced when $\theta>55^\circ$. Although the anisotropic nature is minimal in $^{168}$Er samples, the modification of density by the non-local term of DDI is still significant when dipoles predominantly align in plane which possibly makes local $\widetilde{g}_{eff}$ fail to reveal the universal scaling around BKT transition point in superfluid regime. On the other hand, we find $\widetilde{g}_{eff}$ describes the transition point well and the universality scaled by $\widetilde{g}_{eff}$ survives from normal gas to the superfluid transition point within the experimental resolution. In the EoS measurement, we average out the anisotropic effect due to the minimal density response to the trap along two axes. However, we expect to observe anisotropic compressibility along the x and y directions with improved optical resolution. To unveil the anisotropic behavior of DDI, we focus on local atom number fluctuation in center of the cloud which is in the deep superfluid regime.

\begin{figure*}
\includegraphics[scale=0.52]{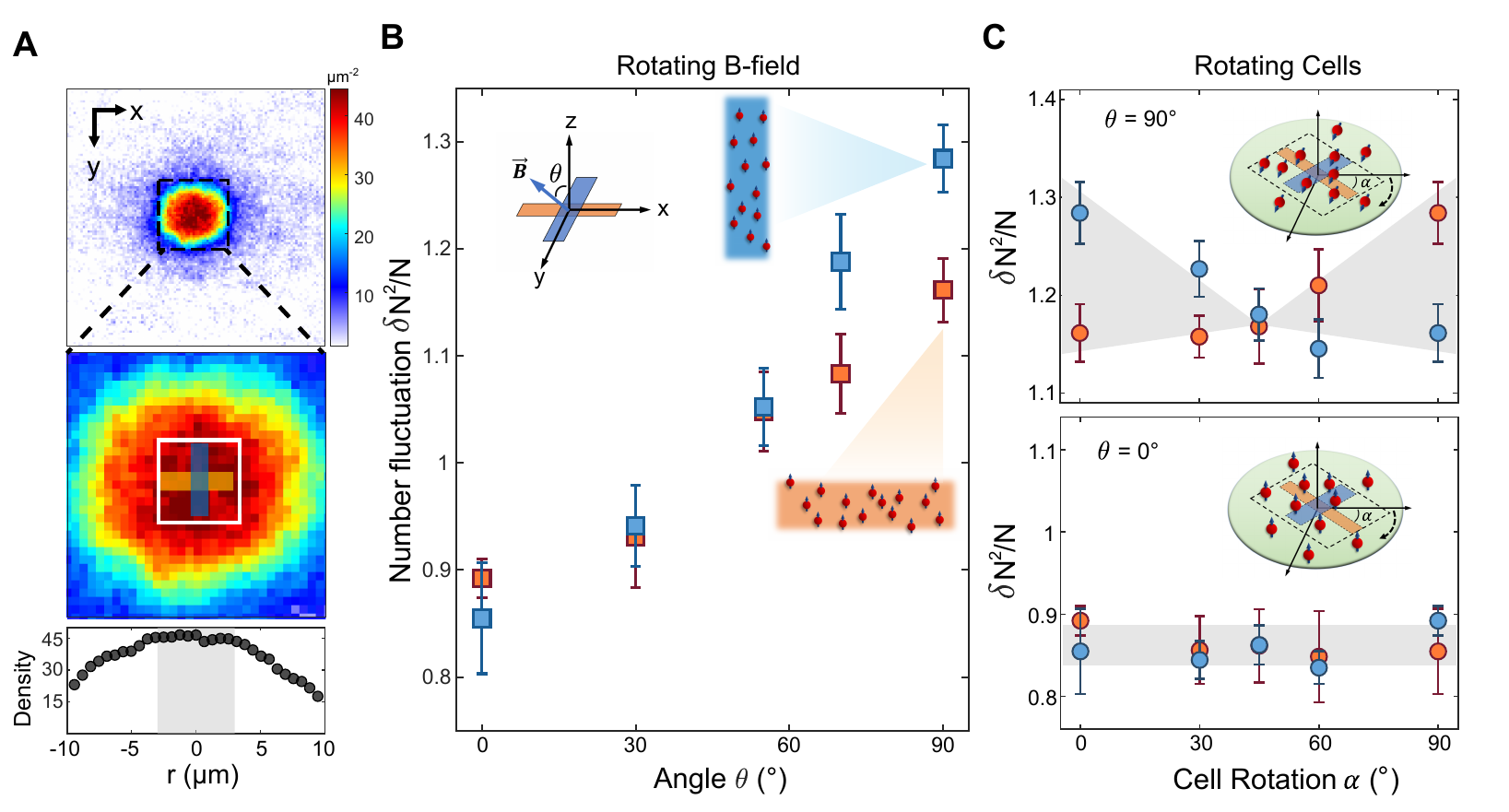}
\caption{\textbf{Anisotropic atom number fluctuation.} \textbf{(A)} Schematic of measuring anisotropic atom number fluctuation. We select two orthogonal rectangular cells within the central, nearly homogeneous region (the white box) of the 2D cloud. We then measure the fluctuation in the number of atoms inside these cells. To probe the number fluctuation, we take all possible configurations of blue (orange) cells inside the central box. Bottom plot is the center cut along the $x$ axis of the cloud for $\theta=90^\circ$. \textbf{(B)} Number fluctuation in the blue cell (blue squares) and orange cell (orange squares) under different tilting angles $\theta$ of the dipoles. Anisotropic number fluctuation emerges when tilting dipoles from $\theta=0^{\circ}$ to $\theta=90^{\circ}$. \textbf{(C)} Rotating two detection cells in a plane, it is clear from the anisotropic $\theta=90^{\circ}$ samples that the fluctuation is enhanced when the rectangular detection cells align with the dipoles. On the other hand, the isotropic $\theta=0^{\circ}$ samples do not exhibit significant variance when two detection cells are rotated. The shaded region serves as guidance for eyes. Error bars denote statistical standard error.}
\end{figure*}

\subsection*{Anisotropic number fluctuation}

Although in a weakly interacting 2D dipolar Bose gas, our measurement suggests the BKT transition point is determined by a local isotropic $\widetilde{g}_{eff}$ similar to contact gases, unique anisotropic features still exist in a 2D dipolar superfluid with tilted dipoles due to the Bogoliubov excitation spectrum $E_{\textbf{k}}=\sqrt{\frac{\hbar^2k^2}{2m}\left[\frac{\hbar^2k^2}{2m}+2nV(\textbf{k})\right]}$ with anisotropic momentum-dependent interaction in 2D $V(\textbf{k})=g_{eff}-3g_{dd}G(kl_z/\sqrt{2})[\cos^2(\theta)-(k_y/k)^2\sin^2(\theta)]$ where $G(q)\equiv \sqrt{\pi}qe^{q^2} {\rm erfc} (q)$ with erfc the comlementary error function, assuming dipoles are tilted toward $y$ axis. When $\theta=0^\circ$, the interaction in momentum space is again isotropic while when $\theta\neq0^\circ$ the anisotropic $V(\textbf{k})$ leads to anisotropic structure factor $S(\textbf{k})$ and density-density correlation $g_2(\textbf{r})$. Such effect can be observed by probing atom number fluctuation in two orthogonal anisotropic cells\cite{baillie2014number}.

To directly probe anisotropic number fluctuations, we utilize two rectangular cells with identical geometry but orthogonal orientations (blue and orange cells in Fig.~4A). These cells are positioned at the center, specifically in the deep superfluid regime. Both cells are located in the nearly homogeneous region at the center of the cloud (the white box in Fig.~4A) wherein the maximum density variance is smaller than 10$\%$~(bottom panel of Fig.~4A), ensuring that all the cells measure the same mean atom numbers. The size of each cell, approximately $1.3\ \mu m\times5.7\mu m$, is larger than both the optical resolution and the healing length $\xi$. Consequently, the phonon mode dominates the fluctuation. We reproduce the sample at each condition around 120 times and obtain the number fluctuation $\delta N^2=\left<N^2\right>-\left<N\right>^2$ where $N$ is the atom number in each cell. We consider fluctuation $\delta N^2/N$ in one cell as one measurement and measure all possible configurations of blue and orange cells within the central box, which results in the mean value of variance and statistical errors.

We first measure $\delta N^2/N$ with different dipole orientations. The result shows that when dipoles are predominantly tilted in plane, the number fluctuation in cells along dipoles direction (blue cell) will be significantly larger than the cell perpendicular to the dipole direction (orange cell) as shown in Fig.~4B which agrees with the prediction in~\cite{baillie2014number}.
We further demonstrate the anisotropic behavior of 2D dipolar samples with dipoles being aligned in plane ($\theta=90^{\circ}$) by rotating the detection cells. As shown in Fig.~4C, when we gradually rotate the detection cells by $90^\circ$, the behavior of fluctuation in two orthogonal cells gradually reverses, particularly at $45^{\circ}$ the difference between two cells disappears. For comparison, the isotropic samples that all dipoles are aligned perpendicular to the plane ($\theta=0^{\circ}$) don't have such anisotropic behavior. 

\subsection*{Conclusion and outlook}
In conclusion, we have realized dipolar superfluid with erbium atoms in a single-layer quasi-2D trap. We experimentally demonstrate the effective local interaction $\widetilde{g}_{eff}$ determines the BKT transition point in quasi-2D dipolar gases with moderate DDI. DDI plays a role like a contact interaction in BKT transition and the strength can be tuned by  changing dipole orientation with respect to the normal direction of the 2D plane, which agrees with the path-integral Monte Carlo simulation for weakly interacting pure 2D dipolar gases~\cite{bombin2019berezinskii,filinov2010berezinskii}. By measuring the EoS of dipolar quasi-2D gases based on LDA, we confirm the scale-invariant behavior of EoS in weakly interacting quasi-2D dipolar gases. However, the universality around BKT transition point described by $\widetilde{g}_{eff}$ is not observed on the superfluid side which can be attributed to the non-local effect in a harmonic trap~\cite{cai2010mean}. Finally, we observe the anisotropic atom number fluctuation in the central superfluid regime of the cloud with dipoles predominantly tilted in plane, which establishes the unique anisotropic correlation in dipolar 2D superfluids~\cite{baillie2014number}. 

Because the non-local term of DDI arises from density variation~\cite{cai2010mean}, it would be interesting to investigate the EoS in a homogeneous dipolar quasi-2D gas~\cite{zhang2021transition} and determine if the universal behavior around BKT transition point is recovered for all dipole orientations. Although in our experiment, the BKT transition point is well described by the short-range picture, it is predicted that anisotropic and long-range interaction between vortices would occur in quasi-2D dipolar gases with larger $a_{dd}/a_s$ or negative $a_s$~\cite{mulkerin2013anisotropic}. This prediction may introduce a new aspect in the dynamics of vortices pairing. It would be possible to study such regime with $^{166}$Er isotope, dysprosium atoms, or microwave-shielded molecules~\cite{bigagli2023observation}. Furthermore, a better understanding of BKT superfluid transition in a single-layer dipolar system provides opportunities to explore superfluidity in bilayer systems coupled by DDI~\cite{pikovski2010interlayer,lutchyn2010spontaneous}, or to investigate the stripe phase~\cite{macia2012excitations,bombin2019berezinskii} and the crystal phase~\cite{buchler2007strongly} in the strongly interacting 2D dipolar system.

\paragraph*{\bf Acknowledgement} We acknowledge support from the RGC through 16306119, 16302420, 16302821, 16306321, 16306922, 16302123, C6009-20G, N-HKUST636-22, and RFS2122-6S04. 

\renewcommand*{\thefigure}{S\arabic{figure}}
\renewcommand*{\theequation}{S\arabic{equation}}
\setcounter{figure}{0}
\setcounter{equation}{0}

\newpage

\section*{SUPPLEMENTAL MATERIALS}

\subsection*{I. Experimental details}
\paragraph*{2D optical trap } The 2D confinement is provided by a 532~nm optical sheet beam tightly focused along the z direction. Additional in-plane confinement is set by a vertical optical dipole trap (vODT). We characterize the axial trap frequency by quenching the sheet beam power to excite a dipole mode of center of mass~(COM) motion along axial direction and hold for different time, then measure the COM position after 10~ms time of flight~(TOF) to obtain $\omega_z$. We use a magnetic gradient field to excite the dipole mode of COM motion in both $x$ and $y$ directions simultaneously and measure the COM position in $x-y$ plane from high resolution image system after 7~ms TOF. The frequency and direction of two planar eigen dipole modes are extracted by principle component analysis (PCA), from which we determine $(\omega_x,\omega_y)$ and the $x-y$ axis. The total trap frequency is $(\omega_x,\omega_y,\omega_z)=2\pi\times(14.3,17.0,1070)~Hz$. The optical sheet beam is linearly polarized along the $x$ axis so that the vector polarizability is vanished and tensor polarizability remains constant when magnetic field is rotated in $y-z$ plane. Therefore, the axial trap frequency with different dipoles orientation remains constant. The polarization of vODT is unclear because it passes through some waveplates for other wavelength and the position is not accessible. To directly evaluate the effect on anisotropic polarizability of vODT, we measure the radial trap frequency of atoms with different dipoles orientation ($\theta=0^\circ,55^\circ,90^\circ$) by quenching vODT. To maximize the contribution from vODT to radial trap frequency, we minimize the sheet beam power and increase vODT power by 4 times of typical working value, which results an almost isotropic in plane confinement confirmed by a circular \textit{in situ} cloud with $\theta=0^\circ$. We provide the result of radial frequency $\omega_r$ (when quenching vODT we always only excited one planar dipole mode) of different dipoles angle in Fig.~S1. Such minimal difference ensures that the anisotropic polarizability of vODT doesn't cause visible change in radial trap frequency when rotating magnetic field within our experiment resolution. 

\begin{figure}[!h]
\begin{center}
\includegraphics[scale=0.25]{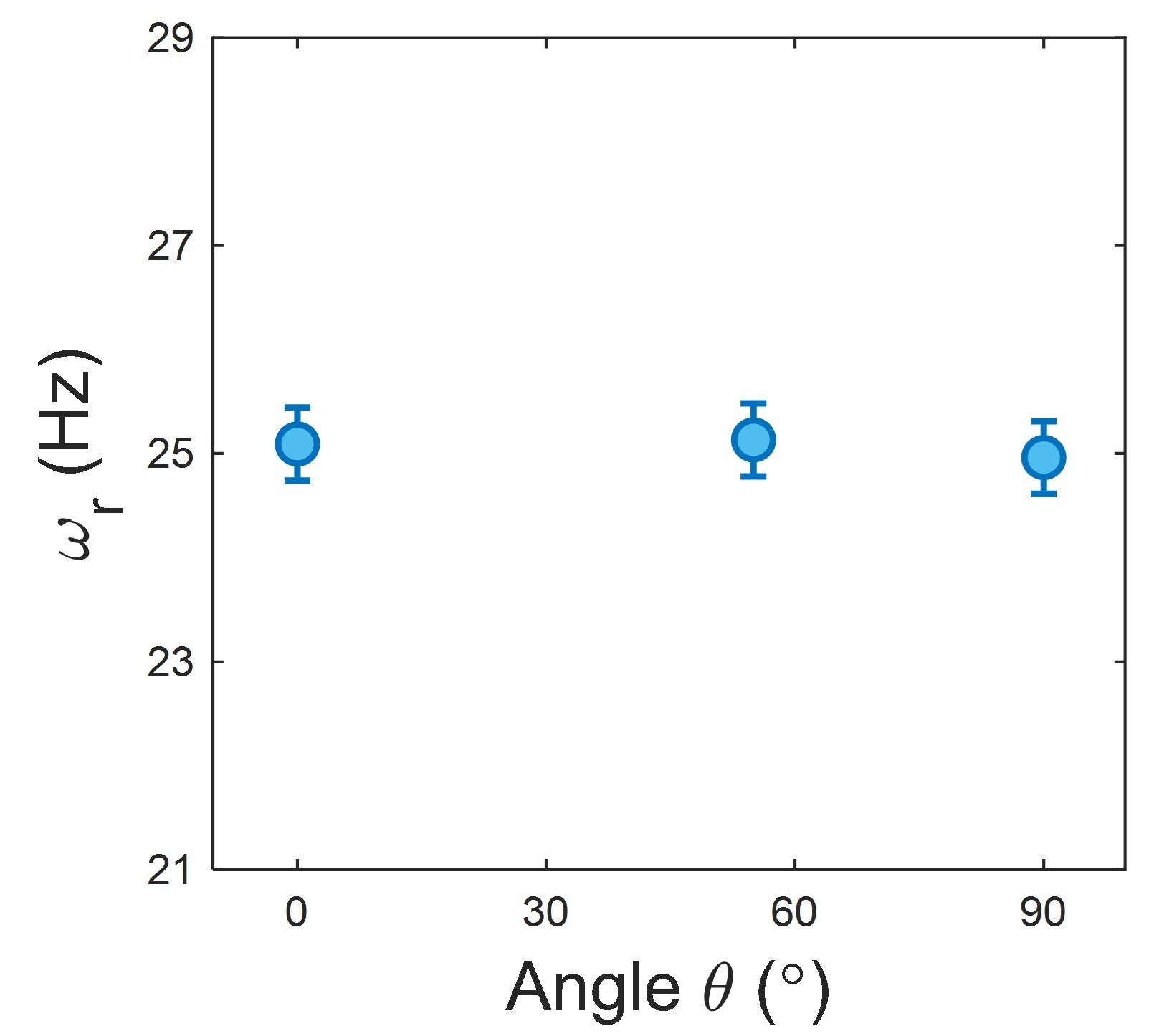}\\
\caption{\textbf{Radial frequency $\omega_r$ of the vODT for different dipole angles.} The polarizability of vODT shows no systematic difference when rotating B-field. Error bars denote 95\% confidence interval of fitting.} \label{S1}
\end{center}
\end{figure}

\paragraph*{Calibration of imaging system}
To precisely determine the critical transition point for the BKT transition, we systematically calibrate the absolute atom number. First, we use the low resolution side image system with a 3D gas in optical dipole trap with frequency $(\omega_x,\omega_y,\omega_z)=2\pi\times(59.5,173.4,183.5) Hz$ to calibrate the atom number. With the condensate fraction $N/N_0$ and temperature $T$ being extracted from the bimodal fitting , the absolute atom number is deduced from the relation $N_0/N=1-(T/T_c)^3$. We take first order correction by considering $T_c=T_c^{ideal}\times(1-3.426a_s/\lambda)$~\cite{smith2011effects}, where $T_c^{ideal}=0.94\hbar\overline{\omega}N^{1/3}/k_B$, $\lambda=h/\sqrt{2\pi mk_BT}$ is the thermal wavelength. The shift of BEC critical temperature by DDI is estimated to be 5\% of the first order correction in our condition~\cite{glaum2007critical} thus is ignored.

We probe {\it in-situ} density distribution with typically $I=2.5I_{sat}\sim5I_{sat}$ using high intensity absorption imaging through the bottom-up high resolution objective lens. The relation between imaging beam intensity and atomic 2D density is given by modified Beer-Lambert law:
\begin{equation}
    \sigma_{eff}n_{2D}(x,y)=-\ln(\frac{I_{out}}{I_{in}})+\frac{I_{in}-I_{out}}{I_{sat}^{eff}}
\end{equation}
where $\sigma_{eff}=\sigma_0/\alpha$ and $I_{sat}^{eff}=\alpha I_{sat}^0$, $\sigma_0=3\lambda^2/2\pi$ and $I_{sat}^0=\pi hc\Gamma/3\lambda^3$ are the light scattering cross section and saturation intensity of an ideal two-level system, $\alpha$ is the correction factor for imperfect polorization and detuning\cite{hueck2017calibrating}. We first calibrate $I_{sat}^{eff}$ for $\theta=0^{\circ}$ by using $I_{in}$ ranging from $0.03I_{sat}$ to $3I_{sat}$ to take image for 2D gases and determine $I_{sat}^{eff}$ by minimizing the variation of atom number given by image taken with different $I_{in}$ \cite{reinaudi2007strong}. For different B-field directions, we calibrate corresponding $\alpha^{\theta}$ for $\theta=0^{\circ}, 30^{\circ}, 55^{\circ}, 70^{\circ}, 90^{\circ}$, respectively. We use low intensity image with $I_{in}=0.03I_{sat}$ to probe the in-situ 2D sample and compare the atom number with the well-calibrated side image system to  extract $\alpha^{\theta}$ for all five angle. The $I_{sat}^{eff}$ for different angle can be deduced by $I_{sat}^{eff0}\times\alpha^{\theta}/\alpha^0$. The pulse duration of image beam is $10\mu s$ and estimated to cause less than $400~nm$ diffusion of atoms which is smaller than the length corresponds to 1 camera pixel.

For momentum focusing imaging, we exploit a separate top-down low resolution imaging system, which provides a larger depth of focus~(DOF). The imaging  pulse duration is $20~\mu s$ with $I=0.5I_{sat}$. The calibration of $\alpha$ and $I_{sat}^{eff}$ is the same as high resolution imaging system.

\paragraph*{Characterize the interaction strength}
The contact interaction strength in a quasi-2D trap is descrbied by $g_c=\sqrt{8\pi}\hbar^2 a_s/ml_z$ where $a_s$ is s-wave scattering length and $l_z=\sqrt{\hbar/m\omega_z}$ is the harmonic oscillator length of axial ground state. During the experiment, B-field is set at 600~mG. The B-field dependent scattering length $a_s$ is modeled as\cite{jachymski2013analytical}
\begin{equation}
    a_s=(a_{bg}+sB)\prod \limits_i(1-\frac{\Delta B_i}{B-B_i})
\end{equation}
where B is the strength of B-field, $i$ denotes $i^{th}$ Feshbach resonances. Using the latest calibration result of $a_{bg}$, $s$, $B_i$ and $\Delta B_i$ for $^{168}$Er\cite{patscheider2022determination}, we obtain $a_s=140a_0$ under 600~mG which exhibits minimal heating away from the Feshbach resonance at 912~mG. The $g_c$ remains constant for all experiments described in this paper.

In addition to the contact interaction, the DDI contributes to the local interaction as $g_{dd}[3\cos^2(\theta)-1]$ where $g_{dd}=\sqrt{8\pi}\hbar^2 a_{dd}/ml_z$,  $a_{dd}=\mu_0\mu^2m/12\pi\hbar^2$, $\mu=7\mu_B$ and $\mu_B$ is Bohr magneton. We tune $g
_{dd}[3\cos^2(\theta)-1]$ in the experiment through changing the B-field direction $\theta$ from z axis. The strength of DDI relative to contact interaction is characterized by $\epsilon_{dd}=a_{dd}/a_s$ which is around 0.47 in our experiments.

The effective local 2D interaction strength is written as
\begin{equation}
    g_{eff}=g_c+g_{dd}(3\cos^2\theta-1)
\end{equation}
and $g_{eff}>0$ is fulfilled for all dipoles orientation in our experiment thus prevents any phonon instability or collapse. We can also define dimensionless effective local 2D interaction strength
\begin{equation}
    \widetilde{g}_{eff}=\frac{\sqrt{8\pi}}{l_z}[a_s+a_{dd}(3\cos^2(\theta)-1)]
\end{equation}

In our experiment, bias B-field is generated by three orthogonal pairs of Helmholtz coils and stabilized by PID feedback circuits. The B-field strength generated by each pair of coil is calibrated by RF spectroscopy. The dipoles are polarized by the B-field so we can tune the dipoles polar angle $\theta$ by changing the B-field direction. The uncertainty of B-field strength calibration is estimated to be 10~mG for each direction, which may cause $\pm1a_0$ uncertainty on $a_s$ and $\pm1^{\circ}$ uncertainty on angle $\theta$.

\paragraph*{Momentum focusing}
The momentum focusing technique is critical to probe the planar momentum distribution $n(\textbf{k})$ of 2D atomic gases \cite{fletcher2015connecting}\cite{murthy2015observation}. One can find the information of extended coherence in the system by the emergence of the zero momentum peak\cite{fletcher2015connecting}, or extract trap averaged first order correlation function $g_1(r)$ from the fourier transform of $n(\textbf{k})$\cite{murthy2015observation}. Both are exploited to determine the critical point of BKT transition in our experiment. 

The momentum focusing benefits from the phase space evolution in a harmonic trap. By solving the Hamilton's equations with harmonic oscillator Hamiltonian $\hat{H}=\frac{p^2}{2m}+\frac{1}{2}m\omega^2_{foc}x^2$, one gets the solution:
\begin{equation}
\left(
\begin{matrix}
m\omega_{foc}x(t) \\
p(t) \\
\end{matrix}
\right)
=
\left(
\begin{matrix}cos(\omega_{foc}t) & sin(\omega_{foc}t) \\-sin(\omega_{foc}t) & cos(\omega_{foc}t) \\
\end{matrix}
\right)
\left(
\begin{matrix}
m\omega_{foc}x(0) \\
p(0) \\
\end{matrix}
\right)
\end{equation}
so after the system evolves for a quarter period at $t=T/4$ we have the relation
\begin{equation}
    m\omega_{foc}x(T/4)=p(0)
\end{equation}
which makes it possible to extract initial momentum information in the spatial space at $t=T/4$.

The momentum resolution is defined by $\Delta k=m\omega_{foc}\Delta x/\hbar$ where $\omega_{foc}$ is the angular trap frequency of focusing trap and $\Delta x$ is the spatial resolution. In our imaging system, the momentum resolution $\Delta k\approx 0.78~\mu m^{-1}$ sets the largest range over which we can probe the coherence length to $L=2\pi/\Delta k\approx8~\mu m$ which is much larger than the thermal De Broglie wave length $\lambda_{db}\approx0.5~\mu m$. In the experiment we release the 2D sample by turning off the optical sheet beam while keep the vODT on to produce a planar 2D harmonic confinement and let atoms free fall for a quarter period. The beam waist of the vODT is $90~um$ much larger than the thermal diameter of the cloud ($33~\mu m$), which minimizes the imperfect momentum focusing caused by inharmonicity. The Rayleigh range of vODT is $48~mm$, much longer than the free falling distance of atoms ($1.25~mm$) and, therefore, the harmonic trapping frequency remains almost constant for atoms in momentum focusing process.

\vspace{20pt}
\subsection*{II. Analysis of momentum distribution for $N_c^{BKT}$ measurement}

\paragraph*{First order correlation function $g_1(\textbf{r})$}
The first order correlation function $g_1(\textbf{r})$ can be deduced from Fourier transform of momentum distribution\cite{murthy2015observation}
\begin{equation}
    g_1(\textbf{r})=\int{n(\textbf{k})e^{i\textbf{k}\cdot\textbf{r}}d^2k}
\end{equation}
We note that the $g_1(\textbf{r})$ obtained from equation above is trap averaged one which entails the off-diagonal correlations of all particles in trap. We perform azimuthal averaging on $g_1(\textbf{r})$ and obtain a distance-dependent correlation function $g_1(r)$. We observe an algebraic decay that $g_1(r)\propto r^{-\eta}$ when the atom number is over $N_c^{BKT}$ which establishes the extended coherence over the sample and accordingly the existence of BKT superfluid of dipolar gases. On the other hand, the correlation function shows exponential decay $g_1(r)\propto e^{-r/\xi}$ below critical atom number . We further investigate the $g_1(r)$ of the data shown in Fig.~2, and evaluate $\chi^2_{exp}$ for exponential fitting and $\chi^2_{alg}$ for power law fitting of sample with different atom number. From Fig.~S2 we can observe BKT transition happens at the position where $\chi^2_{alg}$ exceeds $\chi^2_{exp}$ and the transition point closely agrees with the result obtained by zero momentum peak in Fig.~2. Here the atom number is the raw atom number by counting all atoms in momentum focusing images without subtracting atoms in axial excited states.

Because the decay exponent $\eta$ measured by power-law fitting is a trap average value, it doesn't recover to ideal value $\eta=0.25$ at transition point as in homogeneous case\cite{murthy2015observation}.

\begin{figure}[!h]
\begin{center}
\includegraphics[scale=0.3]{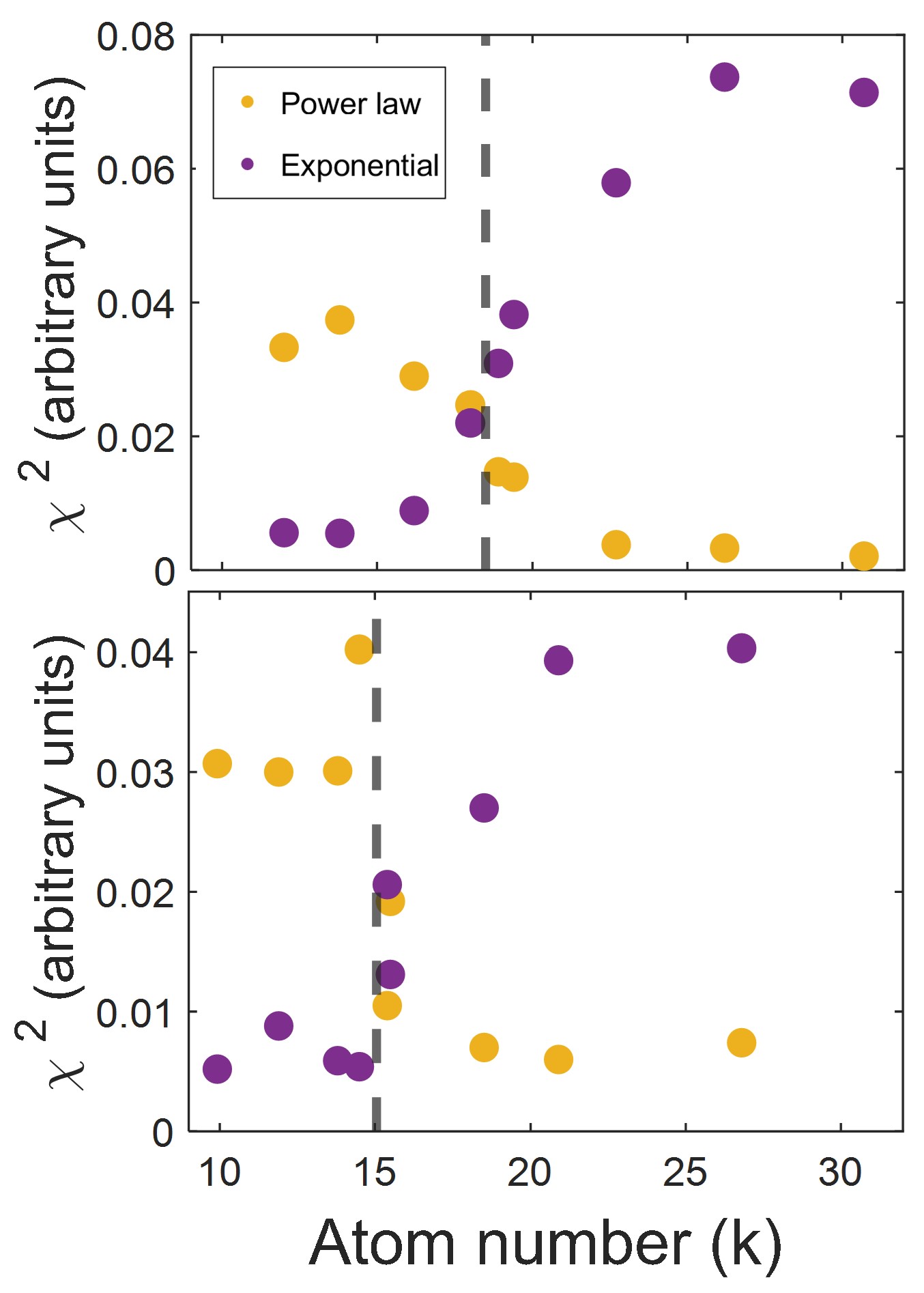}\\
\caption{\textbf{$\chi^2$ for exponential and algebraic fitting} of $g_1(r)$. Upper (Lower) panel is for $\theta=0^\circ(90^\circ)$ of the same dataset used in Fig.~2. Yellow dots represent algebraic fitting and purple dots represent exponential fitting. Dash line shows the raw $N_c^{BKT}$ measured in Fig.~2.} \label{S3}
\end{center}
\end{figure}

\paragraph*{Thermometry in momentum space}
Our thermometry for momentum distribution follows the method in\cite{fletcher2015connecting} within a mean-field framework. Because the axial trap frequency $\omega_z$ and planar trap frequency $\omega_x, \omega_y$ are significantly different, we assume the wave function $\ket{\psi}$ is locally separable into planar momentum state $\ket{\textbf{k}}$ and axial oscillator states $\ket{\phi_j}$
\begin{equation}
    \ket{\psi}=\ket{\textbf{k}}\ket{\phi_j}
\end{equation}
Under  the local density approximation(LDA), a local chemical potential is given as $\mu_L(\textbf{r})=\mu-V(\textbf{r})$ where $V(\textbf{r})$ is the planar trapping potential. We use a semi-classical model that treats the axial motion $\ket{\phi_j}$ as discrete quantized states while the planar motion $\ket{\textbf{k}}$ as continuous states. The momentum distribution could be found by the integrating Bose-Einstein distribution in position space
\begin{equation}
    n(k)=\sum_j\int d^2\textbf{r}\left[e^{\beta(\frac{\hbar^2k^2}{2m}+\epsilon_j-\mu_L(\textbf{r}))} \right]^{-1}
\end{equation}
For a non-interacting system the the axial energy would simply be $\epsilon_j=j\hbar\omega_z$, but should be modified by the presence of interaction. We first evaluate the set $\epsilon_j(\mu_L)$ as a function of local chemical potential by perturbatively solving the axial Schr\"{o}dinger equation
\begin{equation}
    \left[\frac{-\hbar^2}{2m}\frac{\partial^2}{\partial z^2}+\frac{m\omega_z^2z^2}{2}-\frac{\hbar\omega_z}{2}+2gn_{3D}(z)\right]\phi_j(z)=\epsilon_j\phi_j(z)
\end{equation}
in the basis of the non-interacting axial harmonic oscillator states\cite{fletcher2015connecting}. When solving axial Schr\"{o}dinger equation, we simply treat DDI as a local interaction by considering $g=4\pi\hbar^2[a_s+a_{dd}(3\cos^2\theta-1)]/m$. Then we fit the azimuthally averaged momentum distribution $n(k)$ with the pre-solved $\epsilon_j$ and eq.(9) to extract the temperature and atom number in axial ground state. Because beyond mean-field correlations mainly affect the highly populated low-k states\cite{holzmann2010universal}, we restrict our fitting to high momentum tail $p^2/2mk_BT\gtrsim \widetilde{g}$ where the occupation of momentum states hasn't been significantly modified by beyond mean-field correlation.

We compare the temperature extracted from momentum focusing image with the one processed with traditional gaussian fitting from side imaging after 16~ms TOF. We produce samples with variable hold time and $\theta$ and extract the temperature in Fig.~S3. In general, average temperature given by two thermometry agrees with each other within $1.5\ nK$ although the temperature given by fitting from side imaging is slightly lower than  the one from momentum focusing. 

In the dataset used for Fig.~2, the temperature and atom population in the axial ground state are $54.5\pm1~nK$ and $76\pm2\%$, respectively, for all orientations $\theta$. This supports our assumption that the population of the ground state and temperature are not influenced by the orientation of the dipoles. However, the absolute $N_c^{BKT}$ that only counts the atom number in the axial ground states is $10\%$ lower than the prediction given by
 
\begin{equation}
   N_c^{BKT}=\left[1+3\frac{\widetilde{g}_{eff}}{\pi^3}\ln^2{(\frac{\widetilde{g}_{eff}}{16})}+\frac{6\widetilde{g}_{eff}}{16\pi^2}[15+\ln{(\frac{\widetilde{g}_{eff}}{16})}]\right]
\end{equation}
with $N_0=\frac{\pi^2}{6}(\frac{k_BT}{\hbar\omega_r})^2$ and temperature $T$ directly fitted from momentum distribution for all five angle $\theta$ we measured. We attribute this overall shift to the systematic error on calibration of the atom number and imperfect thermometry. Note the critical atom number is sensitive to the temperature as $N_c^{BKT}\propto T^2$ in a 2D harmonic trap.

\begin{figure}[!h]
\begin{center}
\includegraphics[scale=0.2]{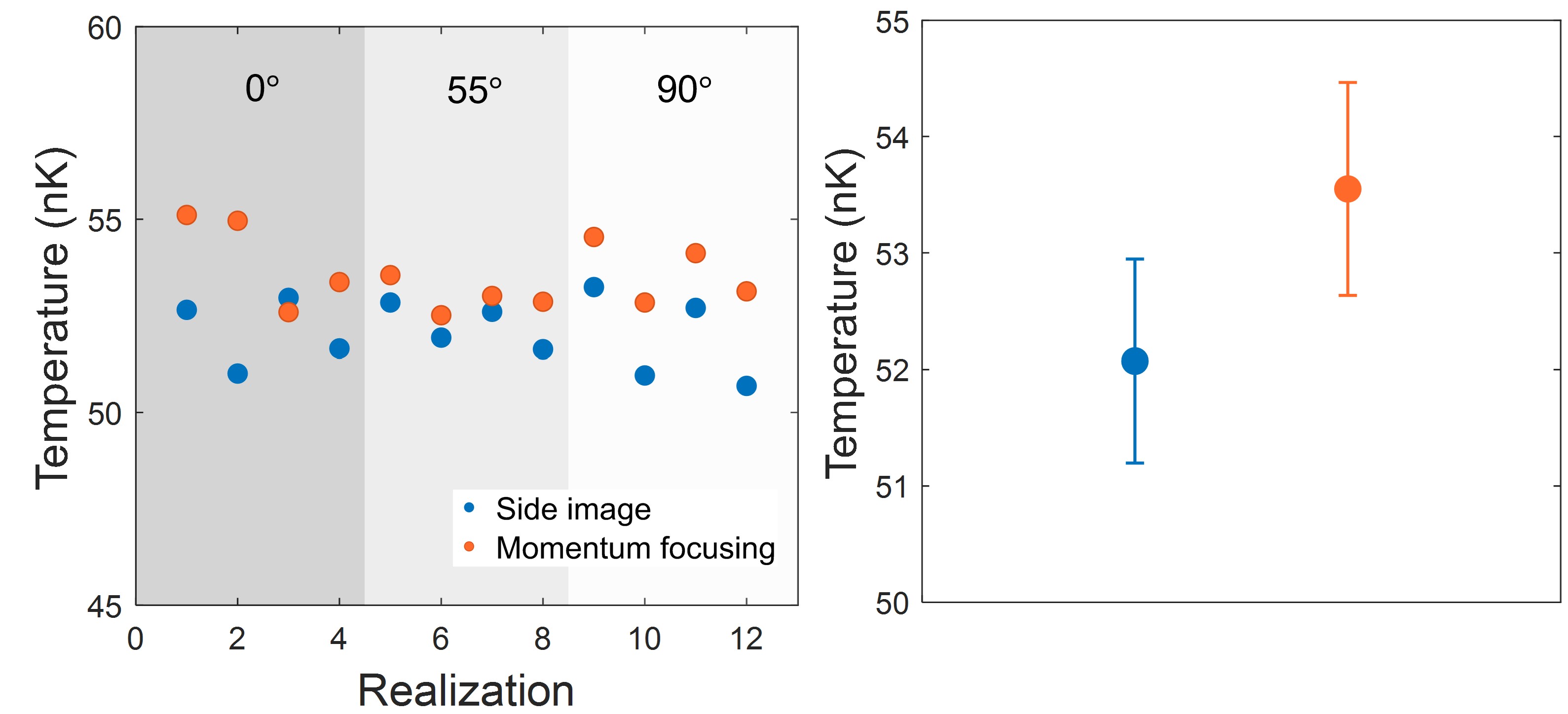}\\
\caption{\textbf{Comparision of thermometry for a 2D sample}.  \textbf{Left}: Extract temperature from momentum focusing image~(orange filled circles) and side image~(blue filled circles) for samples with different atom number and $\theta$ around transition point. \textbf{Right}: Mean value of fitted temperature in left panel. Error bars are standard deviation of the measurement. Blue~(Orange) filled circle is for side~(momentum focusing) image.} \label{S3}
\end{center}
\end{figure}

\paragraph*{Error analysis of $N_c^{BKT}$ measurement}
We first use a quartic polynomial to fit the data $N\ vs.\ t$ shown in Fig.~2A upper panel to get $N(t)$ as a continuous function of holding time. Then we use the critical time $t_c$ discussed in main text to determine raw critical number $N^{BKT}_c=N(t_c)$. The error bars present in main figure of Fig.~2B are fully statistical. For critical point measurement of each angle, we have totally around $30$ different holding time for plotting $P_0\ vs.\ t$ to determine the onset of extended coherence. We employ bootstrapping analysis to evaluate the statistical errors in which we randomly remove $15\%$ data points and extract $t_c$ by piecewise linear fitting. By repeating this process 500 times, we calculate the standard deviation of the resulting $t_c$ and use the corresponding $N_c$ as our measure of statistical uncertainty~\cite{fletcher2015connecting}.

There are two primary sources of systematic errors: the absolute atom number calibration and the thermometry. We anticipate an uncertainty of $\pm5\%$ for the absolute atom number. The extraction of temperature, using two different types of thermometry, results in about a $3\%$ difference in temperature leading to a $6\%$ uncertainty in the predicted $N_c^{BKT}$. This two parts together contribute a systematic uncertainty comparable to the overall shift of our measurement on absolute $N_c^{BKT}$. The systematic uncertainty is presented as shaded region in the inset of Fig.~2.

The observation of relative change on $N_c^{BKT}$ when changing dipoles orientation $\theta$ is not affected by systematic uncertainty coming from calibration and model-dependent fitting.

\begin{figure*}
\begin{center}
\includegraphics[scale=0.35]{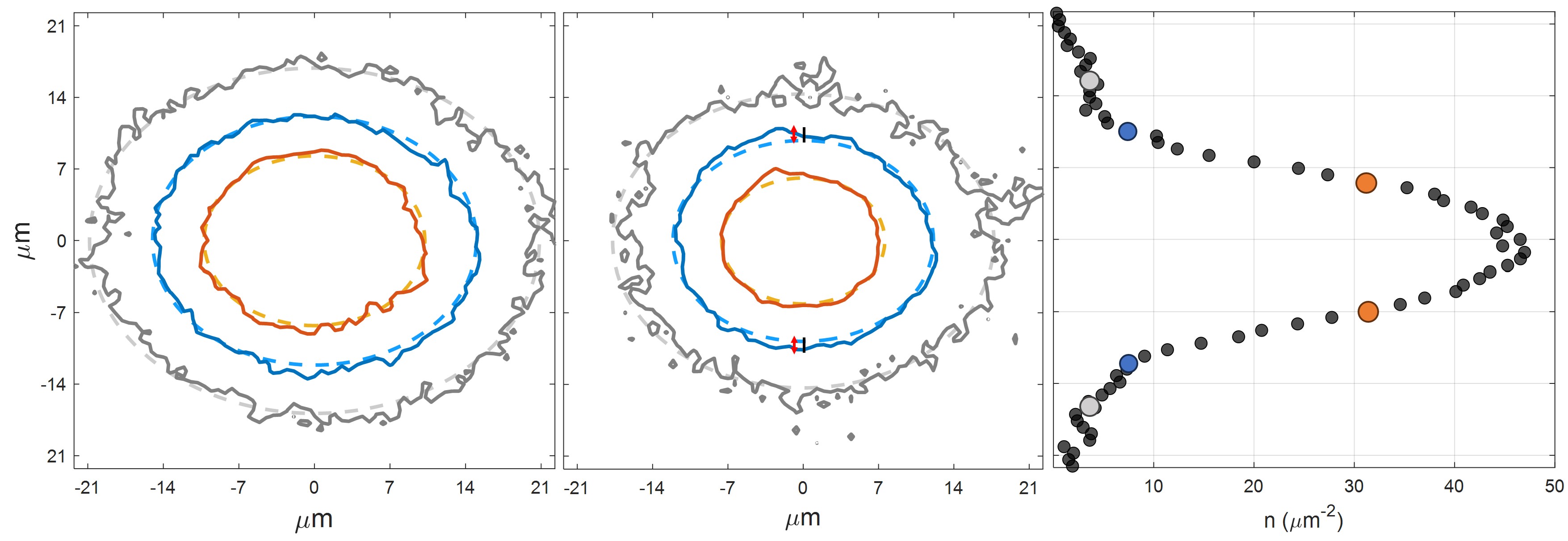}\\
\caption{\textbf{Visualize the small magnetostriction effect in dipolar 2D gases}. \textbf{Left(Middle)}: Equipotential curve~(dash line) and equidensity curve~(solid line) of 2D sample with $\theta=0^{\circ}$~($\theta=90^{\circ}$). Red arrows in the middle are guide for eyes of the deviation between density and trap and black solid line is the optical resolution. \textbf{Right}: Center cut of density profile of $\theta=90^{\circ}$ sample along $y$ axis.} \label{S4}
\end{center}
\end{figure*}

\subsection*{III. Analysis of \textit{in situ} density distribution}
\paragraph*{Azimuthal average based on LDA}

To analyze the {\it in-situ} 2D density distribution $n(x,y)$, we azimuthally average the measured profile over the curve with equal local chemical potential $\mu_L(x,y)=\mu-m\omega_x^2x^2/2-m\omega_y^2y^2/2$ by
\begin{equation}
    n(r)=\left<n(x,y)\right>\bigg|_{{\omega_x^2x^2+\omega_y^2y^2=\omega_r^2r^2}}
\end{equation}.
Here, $n(r)$ represents the averaged density at local chemical potential $\mu_L=\mu-m\omega_r^2r^2/2$ where $\mu$ is the global chemical potential and $\omega_r=\sqrt{\omega_x\omega_y}$. This approach is widely used in the analysis of 2D {\it in-situ} profile of atom with contact interaction~\cite{hung2011observation,yefsah2011exploring,tung2010observation}. Applying this method to the sample with dipoles perpendicular to the plane($\theta=0^{\circ}$) is uncontroversial because in this case DDI is fully isotropic in the plane and the shape of the cloud closely follows the trap geometry~(Fig.~S4 left). However, when dipoles are tilted into the plane, the shape of the cloud no longer perfectly follows the trap due to magnetostriction effect~(Fig.~S4 middle). 

In our system of $^{168}$Er, the strength of DDI ($a_{dd}/a_s=0.47$) is moderate so as to induce the minimal magnetostriction effect. For example, the magnetostriction effect viewed by density profile is minimal as shown in Fig.~S4 middle. Using 2D gaussian fitting, we observe a $6\%$ difference in the aspect ratio between samples with $\theta=0^{\circ}$ and $\theta=90^{\circ}$. This difference is close to the simulation result in the mean-field regime with parameter similar to ours~\cite{cai2010mean}. As shown in Fig.~S4, the deviation between the equi-density and equi-potential curve (red arrow) is smaller or comparable to the optical resolution (black solid line). This deviation is not resolvable with our imaging system. Therefore, averaging along the dashed and solid curve results in less than $0.3\%$ difference which is much smaller than statistical uncertainty. In our experiment, the minimal anisotropic response to the trap allows us to proceed the traditional azimuthal average method. With improved optical resolution, we anticipate observing anisotropic EoS along various directions.

\paragraph*{Thermometry for density distribution} To extract the global chemical potential $\mu$ and temperature $T$ of the 2D sample from {\it in-situ} density profile, we consider thermally excited states along the axial direction based on the Hartree-Fock mean field (HFMF) prediction~\cite{tung2010observation,yefsah2011exploring}
\begin{equation}
     \begin{split}
    n(r)=-\frac{1}{\lambda_{db}^2}\ln(1-e^{(\mu-\frac{1}{2}m\omega_r^2r^2-2g_{eff}n_0)/k_BT})
    \\-\sum_{j>0} \frac{1}{\lambda_{db}^2}\ln(1-e^{(\mu-\frac{1}{2}m\omega_r^2r^2-j\hbar\omega_z)/k_BT})
    \end{split}
\end{equation}
, and fit the low density thermal tail of the azimuthally averaged profile $n(r)$ where $n_j$ is the density profile of $j^{th}$ axial excited state. Here, we ignore any interaction occurring either between or within the axial excite states during the fitting process because they are much weaker than axial oscillator energy $\hbar\omega_z$ in the normal gas regime. An example of fitting the low density thermal wing by HFMF model is in Fig.~S5 left. Finally, we take into account the inter-level interaction by considering local chemical potential of the $j^{th}$ excited state~\cite{tung2010observation}
\begin{equation}
    \mu_j=\mu-\frac{1}{2}m\omega_r^2r^2-j\hbar\omega_z-\sum_{l\neq j}2\left(\frac{4\pi\hbar^2}{m}af_{jl}n_l(r)\right)
\end{equation}
and intra-level interaction
\begin{equation}
    G_j(r)=2(4\pi\hbar^2a/m)f_{jj}n_j
\end{equation}
$f_{jl}$ is the normalized axial wavefunction overlap integrals between axial quantum states $j$ and $l$. Although the density profile of atoms in the axial ground state no longer follows HFMF prediction when atoms are close to the center of the trap, density profiles of atoms in excited states~($j>0)$ are expected to be described by HFMF in the whole range
\begin{equation}
    n_j(r)=-\frac{1}{\lambda_{db}^2}\ln(1-e^{-[G_j(r)-\mu_j(r))]/k_BT})
\end{equation}.
For given $T$ and $\mu$, $n_j(r)$ are determined self-consistently and $n_0(r)$ is obtained by $n_{tot}(r)-\sum_{j>0}n_j(r)$ as shown in Fig.~S5 right. Thus by plotting $(\mu_0-m\omega_r^2r^2/2)/k_BT\ vs.\ n_0\lambda_{db}^2$ we extract the EoS of atoms in the axial ground state. We map DDI to an effective short range interaction by $a=a_s+a_{dd}(3\cos^2\theta-1)$ when evaluating inter-layer and intra-laying coupling as a first-order approximation. We find that changing inter-level interaction strength by the factor of two only alters ground state population by less than $1\%$, which justifies the first-order approximation should be reasonable for evaluating the density profile in the axial ground state.

\begin{figure}
\begin{center}
\includegraphics[scale=0.17]{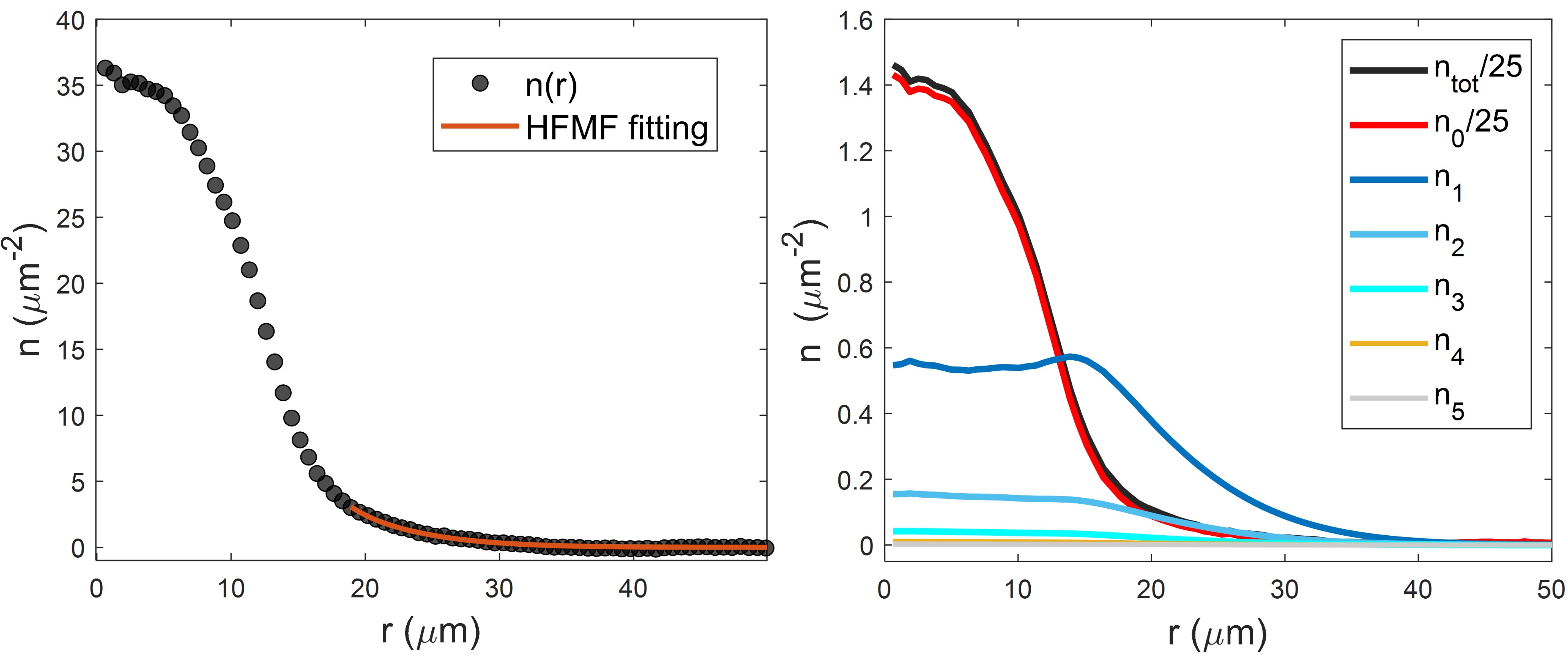}\\
\caption{\textbf{Example of fitting \textit{in-situ} density profile.} \textbf{Left}: An example of using HFMF to fit thermal tail of density profile to extract global chemical potential $\mu$ and temperature $T$. Black dots are azimuthally averaged density profile with dipoles along $z$($\theta=0^{\circ}$), orange curve is HFMF fitting for thermal tail.  \textbf{Right}: Density profile of atoms in axial ground state $n_0$ and in 
 $jth$ excited states $n_j$ extracted from self-consistent fitting. In the superfluid regime the axial ground state population is $>95\%$ while total atoms populate in axial ground state is $>90\%$.}\label{S5}
\end{center}
\end{figure}

\paragraph*{EoS with tilted dipoles}
The profile of quasi-2D dipolar gases is the same as contact gases in the thermal regime, as suggested in \cite{ticknor2012finite}. Therefore, they can be described by HFMF theory $D(\widetilde{\mu})=-\ln[1-\exp(\widetilde{\mu}-D\widetilde{g}_{eff}/\pi)]$~\cite{hadzibabic2008trapped}. In the deep superfluid regime where the system can be describable by classical field theory, we examine the extended Gross-Pitaevskii equation (eGPE) with dipole-dipole interactions (DDI) in a q2D harmonic trapping system~\cite{cai2010mean}. For a pure dipolar condensate, the BEC wave function separates into
\begin{equation}
    \psi(\textbf{r},t)=e^{-i\omega_zt}\psi_{2D}(x,y,t)\phi(z),
\end{equation}
\begin{equation}
    \phi(z)=\left(\frac{m\omega_z}{\pi\hbar}\right)^{1/4}e^{-m\omega_zz^2/2\hbar}.
\end{equation}
Using the dimensionless rescaling $\textbf{r}\rightarrow \textbf{r}a_r$, $t\rightarrow t/\omega_r$, $\psi_{2D}\rightarrow \psi_{2D}\sqrt{N/a_r^2}$ where $\omega_r$ is radial trap frequency and $a_r=\sqrt{\hbar/m\omega_r}$, the 2D wave function fulfills 2D eGPE:
\begin{widetext} 
\begin{equation}
    i\partial_t\psi_{2D}(x,y,t)=\left\{-\frac{1}{2}\nabla^2+V_{2D}+\\
    \frac{\beta_{2D}}{\sqrt{2\pi\gamma}}[1+\epsilon_{dd}(3\cos^2(\theta)-1]|\psi_{2D}|^2+\Phi_{2D} \right\}\psi_{2D}(x,y,t),
\end{equation}
\begin{equation}
    \Phi_{2D}=-\frac{3\beta_{2D}\epsilon_{dd}}{2}[\partial_{\textbf{n}_r\textbf{n}_r}-\cos^2(\theta)\nabla^2]\int{dx'dy'U_{2D}(x-x',y-y')|\psi_{2D}(x,y,t)|^2}.
\end{equation}
\end{widetext} 
Here we assume an isotropic harmonic trap $V_{2D}=(x^2+y^2)/2$, $\beta_{2D}=4\pi Na_s/a_r$ and $\gamma=\omega_r/\omega_z$. The kernel $U_{2D}$ is given by
\begin{equation}
    U_{2D}(r)=\frac{e^{r^2/4\gamma}}{(2\pi)^{3/2}\sqrt{\gamma}}K_0(r^2/4\gamma)
\end{equation}
where $K_0$ denotes modified a Bessel function of second kind. 

The equation reveals that the eGPE includes a local interaction term which is proportional to $\beta_{2D}[1+\epsilon_{dd}(3\cos^2(\theta)-1)]$, consequently showing the characteristics of an effective contact interaction. The local behavior of DDI is enhanced by increasing axial confinement $1/\gamma$. In addition to the local interaction term, a non-local interaction term $\Phi_{2D}$ plays an important role in the eGPE, which is generated by variations in the density but vanishes for a homogeneous system. The numerical simulation in \cite{cai2010mean} shows that the simulated profiles agree with the analytical solution given by Thomas-Fermi~(TF) approximation with only local term for $\theta=0^{\circ}$. Hence, when dipoles align along the $z$ axis, the system behaves similarly to contact gases throughout the entire EoS, as described by $\widetilde{g}_{eff}$. This  indicates the existence of scale invariance in the system as we observe in the experiment. While when dipoles are aligned in plane~($\theta=90^{\circ}$), the simulation result shows the central density is reduced compared with TF approximation due to the non-local term~\cite{cai2010mean}. Such effect is observed in our EoS measurement in the superfluid regime. Despite the non-local term being significant in the condensate regime for $\theta=90^{\circ}$, and scale invariance not being guaranteed, we still observe that all the EoS curves with varying atom number and temperature collapse together, as shown in Fig.~S6. This suggests that the EoS still follows the quasi-2D form $D(\widetilde{\mu})$.

\begin{figure}
\begin{center}
\includegraphics[scale=0.23]{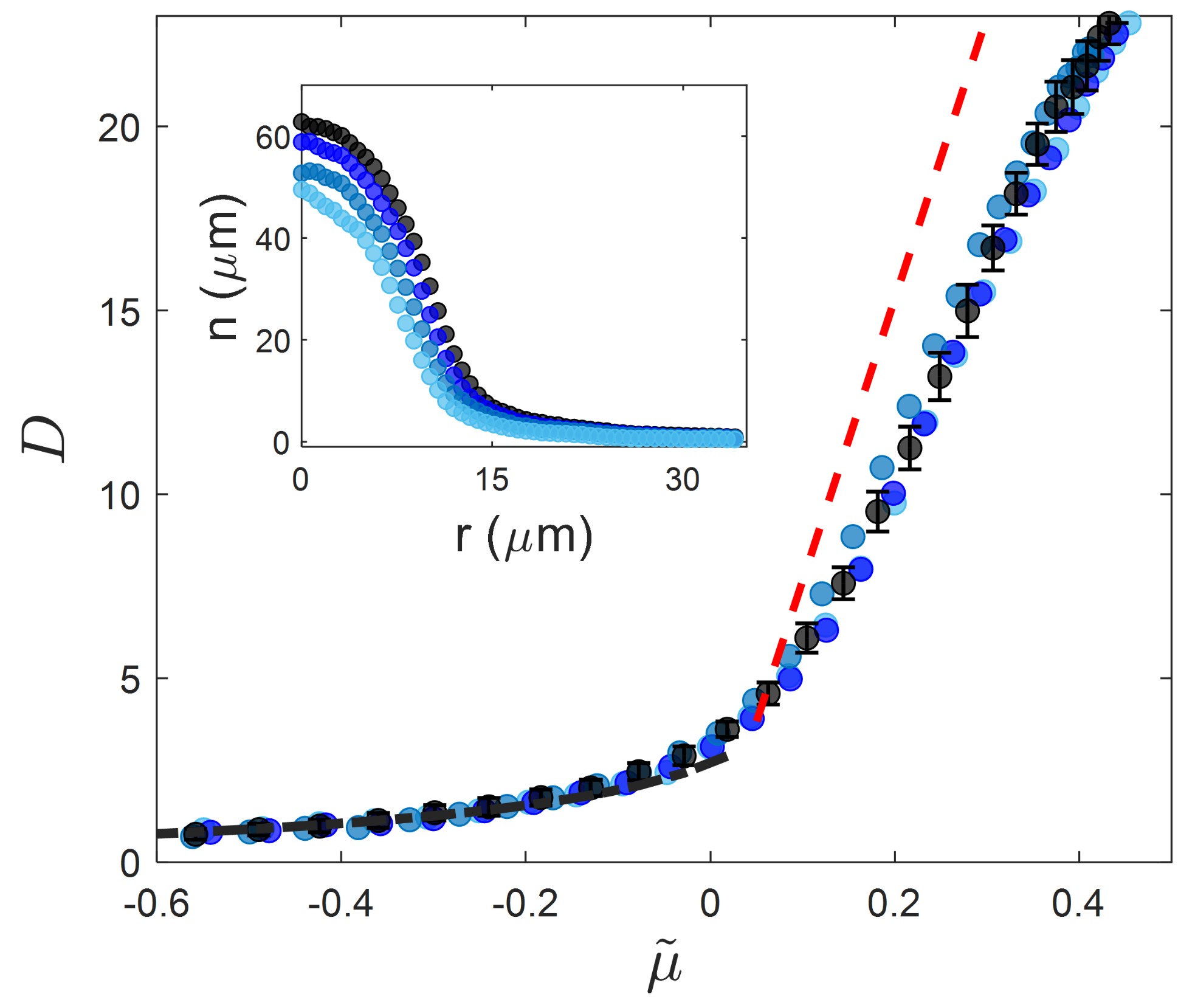}\\
\caption{Scale invariance viewed by EoS of dipolar quasi-2D gases with dipoles in plane~($\theta=90^{\circ}$). The black dash line is HFMF prediction that well describes the profile. The red dash line is TF limit which shows a clear deviation from measurement. Inset is corresponding density profile.}\label{S6}
\end{center}
\end{figure}

\newpage
\paragraph*{Determination of the critical chemical potential from EoS}
After we obtain EoS $D\ vs.\ \widetilde{\mu}$, we compute scaled isothermal compressibility by $\widetilde{\kappa}(\widetilde{\mu})=\partial D/\partial\widetilde{\mu}$. We use function $\widetilde{\kappa}=\widetilde{\kappa}_{sf}e^{y(\widetilde{\mu})}$, $y(\widetilde{\mu})=s(\widetilde{\mu}-\widetilde{\mu}_c)-\sqrt{s^2(\widetilde{\mu}-\widetilde{\mu}_c)^2+w^2}$ to empirically fit the crossover feature of compressibility near transition regime~\cite{hung2011observation}, where the critical scaled chemical potential $\widetilde{\mu}_c$, the compressibility in superfluid regime $\widetilde{\kappa}_{sf}$, the slope of exponential rise $s$ and the width of the transition region $w$ are fitting parameters. The $\widetilde{\kappa}_{sf}$ fitted from the function above is almost equivalent to the slope given by linear fitting on the EoS in the superfluid regime.

\end{document}